\documentclass[twocolumn,showpacs,preprintnumbers,amsmath,amssymb]{revtex4}

\usepackage{graphicx}
\usepackage{dcolumn}
\usepackage{bm}

\begin{document}


\title{A doping dependent specific heat study of the superconducting gap in Ba(Fe$_{1-x}$Co$_{x}$)$_{2}$As$_{2}$}

\author{K. Gofryk$^{1}$}
\email{gofryk@lanl.gov}
\author{A. S. Sefat$^{2}$}
\author{M. A. McGuire$^{2}$}
\author{B. C. Sales$^{2}$}
\author{D. Mandrus$^{2}$}
\author{J. D. Thompson$^{1}$}
\author{E. D. Bauer$^{1}$}
\author{F. Ronning$^{1}$}
\email{fronning@lanl.gov}
\affiliation{$^{1}$Condensed Matter and Thermal Physics, Los Alamos National Laboratory, Los Alamos, New Mexico 87545, USA\\
$^{2}$Materials Science and Technology Division, Oak Ridge National Laboratory, Oak Ridge, Tennessee 37831, USA}


\begin{abstract}

We report a doping, magnetic field and low-temperature dependent study of the specific heat of the iron-arsenide Ba(Fe$_{1-x}$Co$_{x}$)$_{2}$As$_{2}$ at under (x=0.045), optimal (x=0.08) and overdoped (x=0.103 and 0.105) regimes. By subtracting the lattice specific heat the temperature and magnetic field dependence of the electronic specific heat has been studied. The temperature and field dependencies of the superconducting part of $C_{p}$ exhibit similar behavior for all doping concentrations. The temperature variation of the electronic specific heat as well as its field dependence cannot be described by a single isotropic s-wave gap, pointing to a complex gap structure in the system. The lack of doping dependence indicates that the gap structure does not change significantly as a function of doping. We also observe a significant residual linear term of unknown origin in the specific heat of Ba(Fe$_{1-x}$Co$_{x}$)$_{2}$As$_{2}$ which suggests that inhomogeneity may be an important factor in Co-doped BaFe$_{2}$As$_{2}$.

\end{abstract}

\maketitle

The discovery of superconductivity in FeAs-based $R$FeAsO\cite{Kamihara,chen} ($R$-rare earth) has opened a new era in superconductivity studies. Shortly after, other types of superconducting materials containing FeAs layers were discovered including: binary chalocogenides Fe$_{1+x}$Se\cite{hsu,11b}, so called $"111"$ compounds LiFeAs or NaFeAs \cite{111a,111b} and $122$-systems $A$Fe$_{2}$As$_{2}$ where $A$ is an alkali element\cite{122a,122b,122c}. Despite a large theoretical\cite{th} and experimental\cite{ex} effort in the newly
discovered Fe-based superconductors\cite{Kamihara,chen},
the nature of the superconductivity in these materials including the
pairing mechanism and the symmetry of the
order parameter remain unknown. Moreover, the experimental results reported so far are often contradictory, not only between
various techniques, but also between different families. The large sample
and doping dependence may favor scenarios where the low energy excitations, possibly nodal, strongly depend on the particular sample being studied and the probe used to investigate them (e.g. Ref.\onlinecite{kuroki,wang,graser}).

Recently, much attention has been focused on the Co-doped BaFe$_{2}$As$_{2}$ family\cite{122a}
due to the large single crystals which can be produced. They also appear to be
more homogeneous than alternative dopings such as K-doped BaFe$_{2}$As$_{2}$\cite{julien}. At optimal Co-doping (x~=~0.08) an isotropic
gap has been postulated by angle resolved photoemission spectroscopy (ARPES)\cite{terashima} and scanning tunnel microscope (STM)\cite{yin}
measurements, while other experiments such as penetration depth\cite{gordon}, $\mu$SR\cite{williams}, NMR\cite{ning}, thermal conductivity\cite{tanatar, dong, machida}, specific heat\cite{mu}, and Raman scattering\cite{muschler} point
to an anisotropic gap scenario. Several of these measurements are
consistent with the so-called $s\pm$ model with a sign reversal of the order parameter between different sheets of the Fermi surface\cite{s1,s2,s3}. Recently, it
has been suggested by low-temperature thermal conductivity studies, that
in the Ba(Fe$_{1-x}$Co$_{x}$)$_{2}$As$_{2}$ system the superconducting gap evolves from uniformly
large everywhere on the
Fermi surface for low doping to having a very small
value somewhere on the Fermi surface for high cobalt
concentration\cite{tanatar}. Raman scattering measurements have been made which
support this conclusion\cite{muschler}. A similar situation has been argued to exist
in P doped FeAs compounds\cite{has}.

In this paper, we present results of our detailed studies of the specific heat of the Ba(Fe$_{1-x}$Co$_{x}$)$_{2}$As$_{2}$ (x~=~0.045, 0.08, 0.103 and 0.105). By subtracting the lattice contribution, we extract the full electronic $T$-dependence for all compositions studied. A relatively large residual specific heat is related to the presence of the non-superconducting fraction in the samples. The temperature and field variations of the superconducting part of $C_{p}$ exhibit common behavior for all doping concentrations indicating an anisotropic gap structure whose gross features are doping
independent over the range investigated here.

Single crystals were grown out of FeAs flux with the typical size of about 2$\times$1.5$\times$0.2 mm$^{3}$\cite{122b}. The samples crystalize as well-formed plates with the [001] direction perpendicular to the plane of the crystals. The doping level was determined by microprobe analysis. The heat capacity was measured down to 400~mK and in magnetic fields up to 9~T using a thermal relaxation method implemented in a Quantum Design PPMS-9 device. All specific heat data measured in field were field cooled. Magnetic susceptibility have been taken in field cooled conditions with a field of 20~Oe applied parallel to the $ab$ plane of the single crystals. Data on all samples of
similar size and shape were normalized by a constant diamagnetization
factor which gave 1/4$\pi$ for x=0.08.

In general, in the FeAs-based superconductors it is challenging, due to the high $H_{c2}$, to obtain the normal state electronic heat capacity in the superconducting regime. In order to evaluate the electronic contribution of the specific heat, we have used a similar approach used previously for the optimal doped compound Ba(Fe$_{0.92}$Co$_{0.08}$)$_{2}$As$_{2}$ (see Ref.\onlinecite{njp,hardy}). We assume that the phonon part of the specific heat is independent of doping and we use the phonon specific heat obtained from the parent compound. BaFe$_{2}$As$_{2}$ shows a SDW transition at about 140~K. Recent inelastic neutron scattering experiments show that, in the ordered state, spin-wave excitations have a large gap of about 10~meV ($\Delta~\approx$~116~K)\cite{matan}. Therefore, below 40~K, $C_{mag}$ is almost negligible\cite{storey} and we separate the lattice contribution to the specific heat of the parent compound as $C_{ph}$~=~$C^{BaFe_{2}As_{2}}$ - $\gamma_{el}^{BaFe_{2}As_{2}}T$ where $\gamma_{el}^{BaFe_{2}As_{2}}$ is the T$\rightarrow$0 intercept of C/T of BaFe$_{2}$As$_{2}$. Thus, the electronic specific heat, at finite doping, is determined by $C_{el}(T)^{Ba(Fe_{1-x}Co_{x})_{2}As_{2}}~=~C_{p}(T)^{Ba(Fe_{1-x}Co_{x})_{2}As_{2}}-C(T)_{ph}$.
Moreover, a small Schottky-like contribution of about 0.5 mJ/mol K$^{2}$ of the total specific heat at 0.5~K has been also subtracted from the data. The obtained temperature dependence of the specific heat for all samples is shown in Fig.1. For the normal state specific heat below $T_{c}$ we assume the form $(C-C_{ph})/T=\gamma_{n}+bT$. The $bT$ term represents a small correction to the normal state specific heat below $T_{c}$ required for the samples not at optimal doping in order to conserve entropy between the normal and superconducting states at $T_{c}$. The so-derived normal state specific heat is shown by dashed lines in Fig.1. We cannot determine whether the necessary correction to the normal state
specific heat implies either that our assumption of a doping independent
phonon contribution is incorrect, that magnetic contributions are
non-negligible at some dopings, that a pseudogap is present, that quantum critical
fluctuations exist, or that some combination of these effects are at play.
However, we emphasize that our conclusions are independent of the
particular form of the normal state specific heat constructed to conserve entropy.

\begin{figure}[t!]
\begin{centering}
\includegraphics[width=0.5\textwidth]{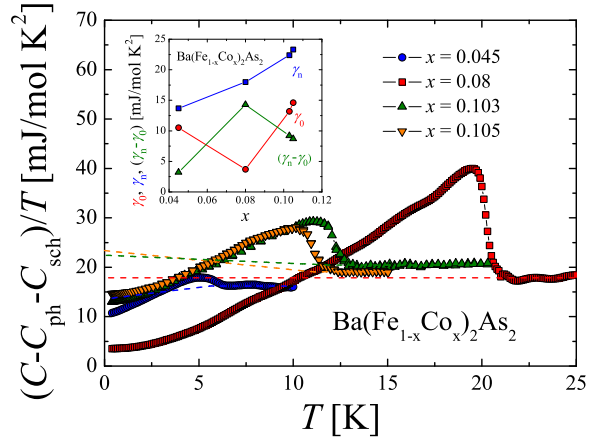}
\caption{(Color online) The low-temperature non-lattice part of the heat capacity of Ba(Fe$_{1-x}$Co$_{x}$)$_{2}$As$_{2}$. The dashed line describes the normal state specific heat (see text). Inset: doping dependence of the residual specific heat ($\gamma_{0}$), normal state specific heat $\gamma_{n}$ and the difference ($\gamma_{n}$-$\gamma_{0}$).}\label{fig1}
\end{centering}
\end{figure}

As can be seen from the Fig.1, at very low temperatures, for all compositions a significant residual specific heat coefficient $\gamma_{0}$ is observed. It ranges from $\gamma_{0}$~=~3.7~mJ/mol~K$^{2}$ for x~=~0.08 to 14.6~mJ/mol~K$^{2}$ for x~=~0.105. The values of the residual specific heat $\gamma_{0}$, normal state specific heat $\gamma_{n}$ and the difference ($\gamma_{n}$-$\gamma_{0}$) as a function of Co concentration are presented in the inset of Fig.1. The smallest value of $\gamma_{0}$ is observed for the concentration close to the optimal Co doping. The residual specific heat strongly increases when moving towards under or overdoped directions. Similar behavior has been previously observed by G. Mu \textit{et.al} in Ba(Fe$_{1-x}$Co$_{x}$)$_{2}$As$_{2}$\cite{mu}. A sizeable value of the low-temperature specific heat has also been reported for Ba$_{0.6}$K$_{0.4}$Fe$_{2}$As$_{2}$ ($\gamma_{0}$~=~7.7~mJ/mol~K$^{2}$)\cite{mu2} and for cuprates superconductors\cite{cuprates,hussey}.

In general, the origin of the residual $\gamma_{0}$ observed in superconducting materials could be caused by pair breaking effects of an unconventional superconductor, crystallographic defects and disorder, and/or spin glass behavior. It is known that for unconventional superconductors the non-magnetic defect and impurities destroy the singularity of the gap at nodes due to breaking the translational symmetry. This process results in a finite density of states induced at the Fermi level. However, the presence of the nodal gap imposed by symmetry, like in cuprates, is ruled out by the vanishingly small residual linear term of the thermal conductivity observed in Ba(Fe$_{1-x}$Co$_{x}$)$_{2}$As$_{2}$\cite{tanatar, dong}. Alternatively, a finite linear term in the specific heat is most commonly identified with regions of the sample which are non-superconducting. To support this simplified notion for these samples, in Fig.2a we plot the non-superconducting fraction of the samples determined in three ways. First, if the superconducting and non-superconducting regions have similar heat capacities then the ratio of $\gamma_{0}$/$\gamma_{n}$ will be equal to the non-superconducting fraction. Additionally, if the superconducting gap structure is unchanged (which we shall demonstrate below) then the condensation energy is simply equal to $A\gamma_{n}T_{c}^{2}$, where $A$ is a property of the superconducting gap structure. Using the normal state specific heat, we extract the condensation energy for all samples by integrating the entropy difference of the normal and superconducting state. In the case of Ba(Fe$_{1-x}$Co$_{x}$)$_{2}$As$_{2}$ this approach gives $U$~=~13.15, 1270, 230 and 180~mJ/mol respectively for x~=~0.045, 0.08, 0.103 and 0.105. If a portion of the sample is non-superconducting, then $U$ will be reduced from its ideal value. Thus by plotting 1-$U/A\gamma_{n}T_{c}^{2}$ we obtain another measure of the non-superconducting fraction. In this case we choose $A$~=~0.22 so that the estimate of non-superconducting fraction by this measure is the same at $x$~=~0.08 as that obtained by $\gamma_{0}$/$\gamma_{n}$. Finally, susceptibility measurements (see Fig.2b) provide the volume of shielded material, which naively should represent the fraction of superconducting material. Consequently, $1-\chi_{(2K)}4\pi$, should again represent the non-superconducting fraction. Within the uncertainty of demagnetization factors and the unknown heat capacity of possible non-superconducting regions, all three methods (see Fig.2a) clearly imply that near optimal doping a majority of the sample is superconducting (~85-100\%), while proceeding to underdoped or overdoped samples a significantly smaller fraction is superconducting (e.g. $<$~25\% for x~=~0.045). Consequently, for the remainder of our analysis we will use the hypothesis that $\gamma_{0}$/$\gamma_{n}$ indeed represents the non-superconducting regions of the sample (most probably due to inhomogeneity), and will discuss its origin in more detail at the end of the paper.

\begin{figure}[t!]
\begin{centering}
\includegraphics[width=0.5\textwidth]{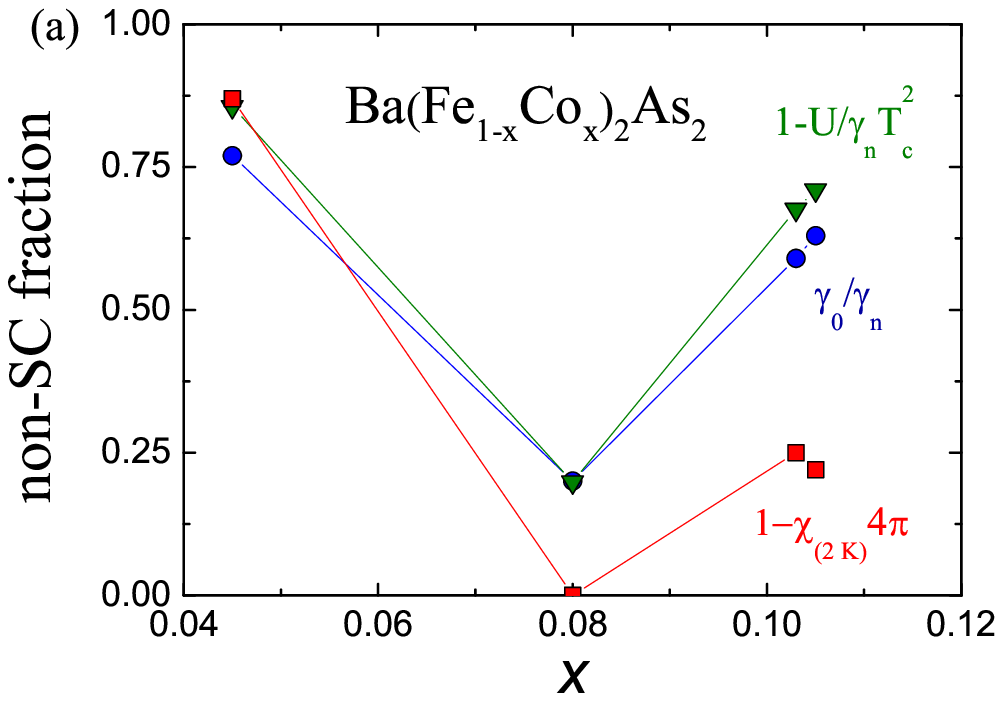}
\includegraphics[width=0.5\textwidth]{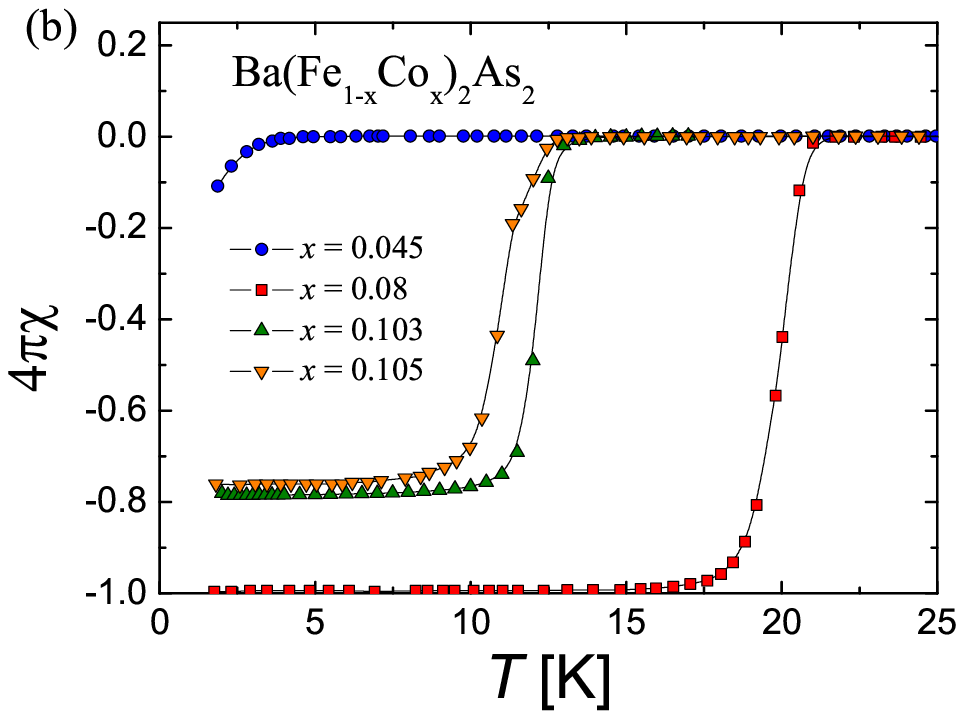}
\caption{(Color online) (a) Doping dependence of the non-superconducting fraction of Ba(Fe$_{1-x}$Co$_{x}$)$_{2}$As$_{2}$ as determined by $\gamma_{0}$/$\gamma_{n}$ (circles), $1-\chi_{(2K)}4\pi$ (squares) and 1-$U/A\gamma_{n}T_{c}^{2}$ (triangles). (b) Temperature dependence of magnetic susceptibility of Ba(Fe$_{1-x}$Co$_{x}$)$_{2}$As$_{2}$. Measurements were made in a field of 20~Oe applied parallel to the $ab$ plane of the single crystals. }\label{fig2ab}
\end{centering}
\end{figure}

Fig.3 displays the electronic part of the specific heat of the superconducting portion of Ba(Fe$_{1-x}$Co$_{x}$)$_{2}$As$_{2}$. It is obtained by subtracting the normal state contribution, together with a small Schottky contribution below 1~K, and normalized by ($\gamma_{n}$-$\gamma_{0}$). Interestingly, taking into account the superconducting fraction of the specific heat only, all the curves collapse below $T/T_{c}$~=~0.7. To evaluate $T_{c}$ we have used the entropy balance shown by the solid red lined in Fig.3. As can be seen from the figure the width of the transition increases away from optimal doping. The specific heat jump $\Delta C/\gamma_{n}T_{c}$~=~1.65 for x~=~0.08 decreases away from optimal doping to 1.5, 1.34 and 1.05 for x~=~0.103, 0.105 and 0.045, respectively\cite{budko}. This observation could result from a small, but noticeable spatial distribution of $T_{c}$ within the crystals. Previous analysis has demonstrated that a single $s$-wave gap cannot reproduce the specific heat data at optimal doping\cite{njp,hardy}. This is illustrated by the single gap fit to the optimally doped sample\cite{njp} shown as the dashed line in Fig.3. Clearly there are additional low energy excitations not captured by a single gap, but can be obtained by a two-gap model all dopings\cite{njp,hardy}.

\begin{figure}[b!]
\begin{centering}
\includegraphics[width=0.5\textwidth]{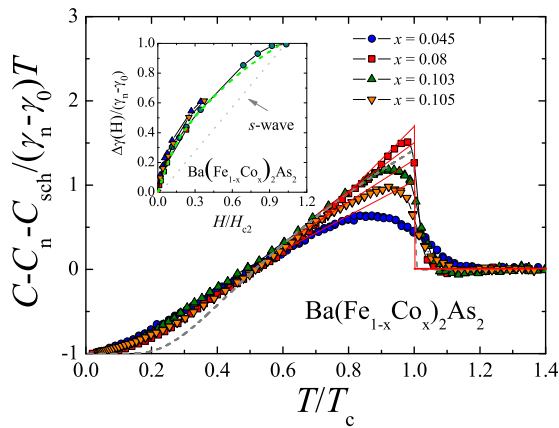}
\caption{(Color online) The normalized temperature dependence of the superconducting state specific heat of Ba(Fe$_{1-x}$Co$_{x}$)$_{2}$As$_{2}$. The dashed grey line represent the specific heat of a single s-wave gap. Inset:  field-induced change in low temperature specific heat obtained at 0 K for H$\|$c by extrapolating the experimental data to zero temperatures (see text). The green dashed line is theoretical curve for $\Delta_{min}/\Delta_{max}$~=~0.5.}\label{fig3}
\end{centering}
\end{figure}

Using a similar approach as presented in Ref.\onlinecite{mu,njp} we have derived $\Delta\gamma(H)=[C(H)-C(0)]/T$ for all samples studied. It is shown in the inset of Fig.3. The data have been presented in the form $\Delta\gamma (H)/(\gamma_{n}-\gamma_{0})$ versus $H/H_{c2}$. The values of $H_{c2}$~=~39, 26, 24 and 8.7~T, respectively for x~=~0.08, 0.103, 0.105 and 0.045, have been obtained from the slope of the upper critical field measured by specific heat and the relation $H_{c2}$=0.69$\frac{dH_{c2}}{dT_{c}}T_{c}$\cite{WHH}. It should be noted that this relation is only a rough approximation strictly valid for a single s-wave gap model. However, values obtained are
reasonable and agree well with those from Ref.\onlinecite{ni}. As can be seen, all curves show roughly the same behavior. It does not match with the behavior expected for a simple $s$-wave order parameter where the localized states in vortex cores induce $\Delta\gamma$ to be proportional to $H/H_{c2}$. The specific heat is changing faster with field than expected for a simple $s$-wave gap scenario. On the other hand, an anisotropic gap will cause the specific heat to deviate from the $H$-linear dependence. A clean $d$-wave superconductor, for example, gives $\Delta\gamma~\propto~\sqrt{H/H_{c2}}$\cite{volovik}. For all dopings we found good agreement with a two gap model\cite{nakai} with a ratio of $\Delta_{min}/\Delta_{max}$~=~0.5 (see the dashed green line in the inset of Fig.3) which is also consistent with the observed temperature dependence. Recently, a similar field dependence of the electronic specific heat has been obtained by Y.~Bang within an $s\pm$ model\cite{bang2} with impurity scattering and a gap size ratio $\Delta_{small}/\Delta_{large}$~=~0.5\cite{bang2}. Our specific heat does not allow us to resolve whether the smaller gap is uniform or possibly even contains accidental nodes. Consequently, we cannot comment on the doping evolution of small changes to the gap structure which may include the lifting of a small nodal component at low energies. However, we can make concrete statements on the lack of doping dependence of the major energy scales. Within a two gap analysis of our specific heat data the ratio of the smaller gap to $T_{c}$ which controls the low temperature and low field properties does not vary by more than 10\% over the doping range studied. The ratio of the larger gap to $T_{c}$, which controls the size of the specific heat jump, may be reduced by as much as 30\% in the underdoped sample and by 15\% in the overdoped samples both relative to optimal doping. However, we cannot rule out the possibility that this small doping dependent evolution of the larger gap is an artifact created by a spread in actual doping concentrations.

These lack of doping dependence to the gap structure is in apparent contradiction with results from Raman and thermal conductivity studies on Co doped BaFe$_{2}$As$_{2}$, which indicate a dramatic evolution of the superconducting gap upon doping\cite{tanatar,muschler}. This may be another manifestation of the extreme sensitivity to sample dependence which appears to be common to the Fe-based pnictide superconductors, although alternative possibilities exist. As alluded to above, the thermal conductivity was measured down to 50~mK, a significantly lower energy scale than the 0.4~K low temperature limit of our measurements. Consequently, the apparent discrepancy could arise from a change in the gap structure below 0.4~K, which we are not sensitive to. Another possibility, which was suggested in a recent P-doped BaFe$_{2}$As$_{2}$ study that encounters a similar contradiction, is that thermal conductivity is more sensitive to light electron pockets which possess the changing near nodal gap structure, while specific heat is more sensitive to the fully gapped bands which possess a higher density of states \cite{kim}. Modeling within an $s\pm$ gap structure is also able to reconcile some of the apparent discrepancy\cite{bang2}. More work is needed to determine whether the Raman results can also be understood in this fashion. Additional doping dependent studies by alternative gap sensitive techniques are required to help resolve these apparent discrepancies.

Before concluding, we return to the origin of the residual linear term of the specific heat. It is tempting to simply attribute the residual linear term in the measured crystals as due to "poor" crystals which possess phase separation on a length scale smaller than that of the EDX probe, and that other measurements were made on better crystals. While acknowledging this possibility, we note that where comparisons are available between these crystals and those measured by other groups grown in different laboratories we are in reasonable quantitative agreement\cite{mu,budko,hardy}. Consequently, we believe the residual linear term and its doping dependence are intrinsic features of the Co-doped BaFe$_{2}$As$_{2}$ system. A residual linear term in the specific heat may be a result of gapless fermionic excitations or from a distribution of two-level systems found in glasses. The magnitude of the residual linear term is too large to be accounted for by a structural glass\cite{hvl}, and a lack of magnetic moments in the overdoped samples\cite{ning} rule out a spin glass origin. Thus, the natural conclusion is that the system is inhomogeneously gapped. The inhomogeneity could exist in real space (phase separation) or momentum space (referring to scenarios where portions of the Fermi surface are ungapped). The latter appears to be ruled out by the lack of a linear term in the thermal transport in zero field\cite{tanatar,dong} as well as by the lack of full diamagnetic shielding across the doping phase diagram (see Fig.3). Macroscopic real space phase separation on the other hand is ruled out by NMR results \cite{julien,ning2}, as well as sample uniformity as probed by our microprobe analysis. However, the NMR lineshape of Ba(Fe$_{1-x}$Co$_{x}$)$_{2}$As$_{2}$ does broaden increasingly as a function of doping\cite{ning2}. Similar behavior is observed in hole doped cuprates and is believed to be a consequence of nanoscaled electronic inhomogeneity\cite{singer}. Strikingly, in cuprates a residual linear term is also observed with a qualitatively similar doping dependence and whose origin is equally perplexing\cite{hussey,wang2}, which could indicate a common origin.  We suggest that the origin of this linear term is connected with nanoscaled electronic inhomogeneity observed in the cuprates, and suggested to be present in the bulk of Ba(Fe$_{1-x}$Co$_{x}$)$_{2}$As$_{2}$ by Ning et al\cite{ning2}. STM measurements of Ba(Fe$_{1-x}$Co$_{x}$)$_{2}$As$_{2}$ confirm the presence of nanoscale electronic inhomogeneity without macroscopic phase separation at the surface\cite{yin}. Further work is required to fully understand the origin of the residual linear term both in the pnictide and cuprate superconductors.

In summary, using the low-temperature specific heat and its magnetic field response, we explore details of the superconducting state in Ba(Fe$_{1-x}$Co$_{x}$)$_{2}$As$_{2}$ at different (under, optimal and overdoped) doping regimes. By subtracting the lattice specific heat the temperature and magnetic field dependence of the electronic specific heat has been studied. The temperature and field dependencies of the superconducting part of electronic specific heat exhibit similar behavior for all doping concentrations. The temperature variation of $C_{el}$ (below $T_{c}$) as well as its field dependence cannot be described by a single isotropic $s$-wave gap, indicating the presence of anisotropic gap structure in the system. Indeed, it has been shown recently for optimally Co-doped BaFe$_{2}$As$_{2}$ samples that a minimum of two superconducting gaps are necessary to describe the temperature dependence of the electronic specific heat\cite{njp,hardy}. Moreover, the lack of doping dependence in Ba(Fe$_{1-x}$Co$_{x}$)$_{2}$As$_{2}$ indicates that the gap structure does not change significantly as a function of doping. The significant residual specific heat observed in this system (see also Ref.\onlinecite{mu,njp,hardy}) is attributed  to a non-superconducting fraction in the sample and suggests that nanoscale inhomogeneity may be an important factor in Co-doped BaFe$_{2}$As$_{2}$.

\begin{acknowledgments}

Work at Los Alamos National Laboratory was
performed under the auspices of the U.S. Department of
Energy, Office of Science and supported in part by the Los Alamos LDRD program. Research at Oak Ridge National Laboratory is sponsored by the Division of Material Sciences and Engineering Office of Basic Energy Sciences.
\end{acknowledgments}

\end{document}